# Centimeter-deep tissue fluorescence microscopic imaging with high signal-to-noise ratio and picomole sensitivity

Bingbing Cheng,[1,2] Venugopal Bandi,[3] Ming-Yuan Wei,[1,2] Yanbo Pei,[1,2] Francis D'Souza,[3] Kytai T. Nguyen,[2,4] Yi Hong,[2,4] Liping Tang,[2,4] and Baohong Yuan[1,2]*

[1]Ultrasound and Optical Imaging Laboratory, Department of Bioengineering, University of Texas at Arlington, Arlington, TX 76010, USA

[2]Joint Biomedical Engineering Program, University of Texas at Arlington and University of Texas Southwestern Medical Center at Dallas, Dallas, TX 75235, USA

[3] Department of Chemistry, University of North Texas, 1155, Union Circle, #305070, Denton, TX 76203, USA

[4]Department of Bioengineering, University of Texas at Arlington, Arlington, TX 76010, USA

*Corresponding author: Baohong Yuan, Mailing address: 500 UTA Blvd, Department of Bioengineering, Arlington, TX 76010, USA, Tel: +1-817-272-2917; FAX: +1-817-272-2251, E-mail: baohong@uta.edu

**Abstract**

Fluorescence microscopic imaging in centimeter-deep tissue has been highly sought-after for many years because much interesting *in vivo* micro-information—such as microcirculation, tumor angiogenesis, and metastasis—may deeply locate in tissue. In this study, for the first time this goal has been achieved in 3-centimeter deep tissue with high signal-to-noise ratio (SNR) and picomole sensitivity under radiation safety thresholds. These results are demonstrated not only in tissue-mimic phantoms but also in actual tissues, such as porcine muscle, *ex vivo* mouse liver, *ex vivo* spleen, and *in vivo* mouse tissue. These results are achieved based on three unique technologies: excellent near infrared ultrasound-switchable fluorescence (USF) contrast agents, a sensitive USF imaging system, and an effective correlation method. Multiplex USF fluorescence imaging is also achieved. It is useful to simultaneously image multiple targets and observe their interactions. This work opens the door for future studies of centimeter-deep tissue fluorescence microscopic imaging.

Fluorescence microscopy has broken many physical limits to allow imaging of biological samples and live tissues with unprecedented resolution, depth, contrast, sensitivity, and specificity[1]. Despite significant progresses, fluorescence microscopy is limited to only thin samples or superficial tissues (~1 mm) because of the extreme difficulty of focusing light into centimeter-deep tissues[1,2]. However, deep tissue fluorescence microscopic imaging is highly desired for at least the following reasons[3-13]. (1) Much interesting *in vivo* micro-information may deeply locate in tissue, such as microcirculation, micro-angiogenesis, micro-lymphangiogenesis, tumor micro-metastasis, and vascularization of implanted tissue scaffolds, etc. (2) Far red or near-infrared (NIR) light can penetrate centimeter-deep tissue via scattering, although the micro-information is completely blurred. (3) Compared with other imaging modalities such as ultrasound (US), computer tomography (CT), magnetic resonance imaging (MRI), or positron emission tomography (PET), fluorescence imaging stands out in the following respects: high sensitivity, low cost, usage of non-ionizing radiation, the feasibility of multi-wavelength imaging, and the feasibility of tissue structural, functional, and molecular imaging.

To conduct microscopic imaging in centimeter-deep tissue via fluorescence, the following fundamental challenges must be addressed[4-13]: (1) how to confine the fluorescence emission into a small volume to achieve high spatial resolution; (2) how to increase fluorescence emission efficiency and sensitively detect those photons to compensate for the signal attenuation caused by the small voxel size in microscopic imaging and by tissue scattering/absorption during the propagation toward the photodetector (i.e., increase signal strength and detection sensitivity to signal photons); (3) how to exclusively differentiate signal photons from other background photons to increase signal-to-noise ratio (SNR) and sensitivity (i.e., increase the detection specificity to signal photons and reduce noise).

To address the first challenge, optical focusing has been replaced by ultrasonic focusing because tissue acoustic scattering is ~1000 times lower than optical scattering[4-12]. Using this method, fluorescence microscopic images with a resolution from microns to hundreds of microns have been demonstrated at millimeters depth in tissue phantoms [4-6,8-10]. Despite these successes, a fundamental barrier faced by these technologies is the quick degradation of SNR and sensitivity to an unacceptable level with the increase of the imaging depth, the spatial resolution, or both. This is mainly due to the quick attenuation of the fluorescence signal below the noise level. Accordingly, centimeter-deep tissue fluorescence microscopic imaging with high SNR and high sensitivity still remains unachievable.

To break this barrier, the above challenges must be addressed. Here, we aim to address them via a recently developed imaging technology, NIR ultrasound-switchable fluorescence (USF)[9]. We discovered and synthesized unique USF contrast agents. The fluorescence emission of these agents can be switched on or off by a focused ultrasound wave[9,10]. The emission intensity on-to-off ratio ($I_{ON}/I_{OFF}$) can reach >200, which is >100 times higher than that of current agent with a similar structure (~1.8)[11,12]. The large value of $I_{ON}/I_{OFF}$ is one of the keys to achieve high SNR and high sensitivity in centimeter-deep tissue. Also, we developed an imaging system to sensitively detect USF signal while significantly suppressing noise. Lastly, we adopted a correlation algorithm to effectively differentiate USF signal from noise. With these unique technologies, we demonstrated for the first time the feasibility of centimeter-deep tissue fluorescence microscopic imaging with high SNR and picomole sensitivity in tissue-mimic phantoms, porcine muscle tissues, *ex vivo* mouse organs (liver and spleen), and *in vivo* mice.

**Results**

**Unique USF contrast agents with $I_{ON}/I_{OFF}$>200 via a new NIR fluorophore.** When encapsulating an environment-sensitive fluorescent dye into a thermo-sensitive nano-capsule, the dye's fluorescence emission exhibits a switch-like function of the temperature (Figure 1(a) and (f))[9,10]. When the temperature is below a certain threshold ($T_{th1}$), the nano-capsule exhibits hydrophilicity and provides a water-rich, polar, and non-viscous microenvironment in which the dye shows very low emission efficiency (so-called OFF). When the temperature is above a threshold ($T_{th2}$), the nano-capsule exhibits hydrophobicity and can dramatically shrink and expel water. Thus, it provides a polymer-rich, non-polar, and viscous microenvironment in which the dye shows strong emission (so-called ON). If the transition bandwidth ($T_{BW}=T_{th2}-T_{th1}$) is narrow, the fluorescence intensity appears a switch function as the temperature.

To achieve high SNR and sensitivity in centimeter-deep tissue, a large value of $I_{ON}/I_{OFF}$ is critical. An extremely environment-sensitive NIR dye was synthesized and characterized (Figure 1(b)–(c)). The dye is an aza-BODIPY (abbreviated as ADP) derivative. The ADP (core) was chemically reacted with two cyanocinnamic acids (CA), and therefore denoted as ADP(CA)$_2$ (see Supplementary Information (SI) for dye synthesis and Figure S1 for the nuclear magnetic resonance (NMR) and mass spectra). ADP(CA)$_2$ has a molecular weight of 927.67 g/M and peak excitation/emission wavelengths of 683/717 nm (in dichloromethane, DCM; Figure 1(c)). They are 27-nm red shifted compared with those of the ADP core (656/690 nm, respectively). Also, the two CAs can provide two carboxyl acid (COOH) groups for conjugation with other units for future use. ADP(CA)$_2$ were found to be extremely sensitive to the polarity (Figure 1(d)) and moderately sensitive to the viscosity (Figure 1(e)) of its solvent. However, it is relatively insensitive to pH and KCl ions in the physiological range (pH: 6.8–7.4; KCl: <150 mM) (Figure S2). Therefore, we expect that it is an excellent candidate for USF imaging.

Commercially available thermo-sensitive polymers (Pluronics) and their co-polymers with polyethylene glycol (PEG) were adopted to synthesize thermo-sensitive nano-capsules for encapsulating the ADP(CA)$_2$: (1) Pluronic-F127, (2) Pluronic-F98, (3) Pluronic-F98~PEG20k; (4) Pluronic-F98~PEG30k; (5) Pluronic-F98~PEG40k. The diameters of these nano-capsules were found to be in a range of 20~70 nm via transmission electron microscopy (TEM) (Figure S3). Their switching relationships of the fluorescence intensity with the temperature are shown in Figure 1(f). Impressively, all the intensity ratios ($I_{ON}/I_{OFF}$) are >200 (Table S1 in SI) while previously developed agents have $I_{ON}/I_{OFF}$ <10[9,11,12,14]. Therefore, we expect that ADP(CA)$_2$-based USF agents should provide high SNR and sensitivity in deep tissue. The final concentrations of ADP(CA)$_2$ and the Pluronic polymers in the injectable agent solution were ~50 μM and ~16.3 mg/mL, respectively. The actual concentrations in the mouse-related studies may be physiologically diluted after intravenous injection. All the above synthesis protocols and measurement methods are provided in SI and Methods.

**Sensitive USF imaging system and effective signal identification algorithm.** Figure 2(a) displays the schematic of the USF system (see the details in SI). Briefly, the excitation laser (671 nm) is modulated at 1 kHz. The fluorescence photons are filtered via three interference filters and two absorptive filters, then refocused on a cooled photomultiplier tube (PMT). The 1 kHz fluorescence signal is finally detected via a lock-in amplifier (LIA). A high-intensity-focused-ultrasound (HIFU) transducer (2.5 MHz) is focused in the sample to heat its focal volume, switch on fluorescence, and then induce the amplitude change of the 1 kHz fluorescence. A motorized translation stage is used to scan the sample. A pulse delay generator (PDG) is used to synchronize different sub-systems.

Figure 2(b)-(c) shows the time sequences of the USF system and the sample setup, respectively. The HIFU exposure time is 300 ms (panel 1) to induce the temperature rise around the focal volume (panel 2). The 1 kHz excitation laser is continuously running (panel 3). The HIFU-induced temperature rise leads to a change in the amplitude of the 1 kHz fluorescence signal (panel 4). After interfering with a phase-locked reference signal (panel 5), the LIA outputs the USF signal (panel 6). The baseline of the LIA output ($I_{BG}$) is subtracted from its maximum value ($I_{Max}$), and the difference ($I_{USF}=I_{Max}-I_{BG}$) is used to represent the USF strength ($I_{USF}$) at that location. $I_{USF}$ can be acquired at each location if the sample or the HIFU focus is scanned. Thus, a USF image can be formed.

The temporal shape of a USF signal does not correlate with background noise (Figure 3(a)–(b)) and is independent of the signal strength (Figure 3(c)). Instead, it is dependent on the type of the agent (Figure 3(d)). Thus, the shape information can be used to differentiate the USF signal from background noise to increase SNR (Figure 3(e)–(f), 5, S4) and also to differentiate one type of agent from another to help multiplex imaging (Figure 4). Specifically, equation (1) shows the correlation coefficient (CrC) between a normalized USF signal at any location (*I(t)*) and a normalized USF reference at a selected location (*R(t)*)[15]. The total data point of *I(t)* or *R(t)* is *N*. If CrC<0.3, it is considered noise, and its $I_{USF}$ is set to zero. If 0.3<CrC<0.8, its $I_{USF}$ is multiplied with the cube of CrC to partially suppress its contribution. If CrC>0.8, no modification is applied on its $I_{USF}$. This correlation method improves the SNR significantly (Figure 3(e)-(f), Figures 4, 5, S4). See details in SI.

$$CrC = \frac{\sum I(t)R(t) - \frac{\sum I(t) \sum R(t)}{N}}{\sqrt{\left(\sum I(t)^2 - \frac{(\sum I(t))^2}{N}\right)\left(\sum R(t)^2 - \frac{(\sum R(t))^2}{N}\right)}} \quad (1)$$

**USF microscopic images in tissue-mimic silicone phantoms and tissue samples**

Five silicone micro-tubes (inner diameter: I.D.=0.31 mm; outer diameter: O.D.=0.64 mm) were inserted through the middle plane (x-y; z=6.5 mm) of a tissue-mimic silicone phantom[16] ($\mu_a$=0.03; $\mu_s'$=3.5 cm$^{-1}$; thickness=13 mm) (Figure 4(a) and Methods). These micro-tubes were filled with different contrast agents (Figure 4(a) and its caption) and imaged using three modalities: directly acquiring fluorescence without ultrasound (DF), USF, and ultrasound (US) (see Methods for details).

The DF images of the tubes are significantly blurred due to the strong light scattering of the phantom (Figure 4(b) and Figure S5). The estimated full-width-at-half-maximum (FWHM) is ~5 mm for the ADP(CA)$_2$-filled tube (Figure S5 and its caption). It is difficult to spatially differentiate the ICG (indocyanine green)-filled tube from the ADP(CA)$_2$-filled tube (Figure 4(b) and Figure S5). In addition, because of the spectrum overlap, both ADP(CA)$_2$- and ICG-filled tubes are seen on the image, no matter which emission filters are used (715-nm long pass (Figure 4(b)) or 711/25 nm band pass (data not shown)).

Figure 4(c) and (d) show the USF images of ICG (color-1) and ADP(CA)$_2$ (color-2), respectively. They are acquired from a single scan using the same system (Ex:671 nm; Em:715 nm long pass), but are differentiated using the correlation method (the spectrum crosstalk is eliminated). Figure 4(e) and (f) show the same USF images overlapped with the DF image. Compared with the DF image, the sizes of the USF images are significantly reduced (FWHM=0.88 mm for ADP(CA)$_2$-filled tubes and 0.95 mm for the ICG-filled tubes) and much closer to the true size of the micro-tubes (also see Figure S5). All the individual micro-tubes can be clearly and correctly resolved.

Figure 4(g) shows the ultrasound images acquired using the same US transducer. The image size (FWHM=0.5 mm, Figure S5) represents the spatial convolution of the acoustic focus of the transducer (lateral FWHM=0.42 mm) with the top inner boundary between the silicon and the liquid (the specific size of the boundary is unknown but should <I.D.=0.31 mm). As expected, the ADP(CA)$_2$- and ICG-filled tubes show much lower acoustic contrast than the air-filled tubes, and they cannot be differentiated from each other because the acoustic wave is insensitive to fluorophore molecules.

Figure 5(a)–(d) show the USF images of micro-tubes that are inserted into porcine muscle tissue of different thickness (0.8, 1.2, 2.2, and 3.1 cm). Figure 5(e)–(h) show the corresponding images processed with the correlation algorithm. All the micro-tubes can be clearly imaged with a similar FWHM (~1.1 mm). Figure 5(i) plots the SNRs as a function of the tissue thickness. Significantly high SNRs have been achieved (17.6–242 after and 18–104 before the algorithm processing). The fitted attenuation coefficient of the SNR versus the tissue thickness (0.756 and 1.049 cm$^{-1}$ before and after the processing) is slightly smaller or close to the average effective optical attenuation coefficient of porcine muscle tissue ($\mu_{eff} \approx 1$ cm$^{-1}$ for 670–900 nm)[17]. This result indicates that (1) tissue's optical attenuation mainly causes the SNR reduction as the thickness increase, and (2) increasing the illumination energy of the light and/or ultrasound in thicker tissue can compensate the tissue optical attenuation (Methods).

Figure 5(j) plots the SNR as a function of the time constant of the LIA (tissue thickness=12 mm). Without applying the correlation algorithm, the SNR increases from 29.9 to 87.9 when the time constant increases from 3 to 300 ms. With the correlation algorithm, the absolute values of SNRs are dramatically improved (between 95.4 and 215.6). The correlation method seems more efficient for short time constants. This is useful to improve USF imaging speed in the future by shortening the time constant without SNR degradation.

**USF microscopic images of *ex vivo* mouse organs and *in vivo* USF signal detection**

The bio-distribution of the USF agent (ADP(CA)$_2$-encapsulated Pluronic-F127 nano-capsules) is shown in Figure 6(a). The left panel is a white-light photo of the major mouse organs, and the right panel is the fluorescence image (Ex/Em:630/700 nm). The stomach, liver, and intestine emit relatively strong fluorescence signal; the spleen and lung show moderate signal; and the heart and kidney show weak signal. A corner of the liver and a part of the spleen were scanned. Figure 6(b) and (c) show the USF microscopic images of liver and spleen on the x-y plane, respectively. The corresponding white-light photos are also shown. The green square (7.62×7.62 mm) in Figure 6(b) and the green rectangle (1.016×15.24 mm) in Figure 6(c) indicate the USF scanning areas. The red and yellow areas in the USF images may represent some blood vessels or microstructures inside the organ that cannot be seen from their white-light photos. Note that the USF images represent a projection of the distribution of the USF contrast agents on the x-y plane along the z-axis (i.e., the ultrasound axis).

The basic *in vivo* setup is shown in Figure 6(d). The HIFU beam is aligned with the photon collection fiber bundle. The focus of the HIFU is positioned inside the mouse body (~5 mm higher from the bottom of the mouse). The mouse thickness at that location is >10 mm. The *in vivo* USF signal was observed and is plotted in Figure 6(e). The open vertical (red) line represents that ultrasound is off, and the three solid vertical (red) lines indicate that three ultrasound pulses are delivered to the animal. Clearly, once an ultrasound pulse is applied, a USF pulse is generated (the blue line). The negative USF pulse relative to the baseline may be caused by the interaction of the USF agents with the *in vivo* biological environment; this phenomenon is discussed in the next section and SI.

**Discussion**

Compared with DF imaging (at centimeters depth), USF improves the spatial resolution >5 times using current setup (2.5 MHz ultrasound). The resolution of the current system is mainly limited by thermal diffusion (see SI). It can be further increased in the future by microscopically heating the contrast agents to achieve thermal confinement (see SI). Once thermal confinement is achieved, adopting an ultrasound transducer with a high frequency and a small f-number can further significantly increase the resolution.

Current USF technology can sensitively and specifically detect USF fluorophores (ADP(CA)$_2$) with a concentration of ~50 μM at a depth of several centimeters using a laser and a HIFU below the safety thresholds (laser intensity: 0.21–3.19 mW/cm$^2$; HIFU's mechanical index: 1.7 when imaging depth <30 mm, and 2.19 when imaging depth >30 mm). Because of the high spatial resolution and the adoption of micro-tubes, the ADP(CA)$_2$-occupied volume in a single voxel in this study is as small as ~68 nanoliters (assuming the volume is a cylinder with a diameter of 0.31 mm (I.D.) and a length of 0.9 mm (the FWHM of the USF image)). Thus, the corresponding number of ADP(CA)$_2$ molecules in this volume is as low as ~3.4 picomoles. As a comparison, conventional fluorescence molecular tomography (DF-based imaging) is difficult to detect <10 picomole molecules of a commercially available dye (AF680) in a mouse phantom at a depth of 11 mm (i.e., a thickness of 22 mm when converting into a transmission geometry)[18].

To push this technology for *in vivo* applications, the following directions should be pursued in future. (1) Optimize current or develop new synthesis strategies for high-threshold ($T_{th1}$>37 °C) USF contrast agents. We did not use them for the mouse-related studies in this study because their synthesis yields and shelf lifetime are low. (2) Identify the mechanisms of the negative USF signal in real biological tissues. (3) Explore more USF fluorophores to cover 750–900 nm NIR spectrum for multiplex imaging and even deeper tissue imaging. (4) Significantly increase the imaging speed, minimize artifacts caused by animal motion, and possibly catch dynamic events. (5) Achieve relatively uniform resolution along both lateral and axial directions for three-dimensional imaging. (6) Further improve system sensitivity by increasing photon-collection efficiency and adopting more efficient technologies to suppress the background photons. (7) Modify the system suitable for scanning a live animal. Currently, it is difficult because of the uneven mouse body, the animal motions, and the limited imaging speed. This is why we only obtained the *in vivo* USF signal instead of an image. Fortunately, all these challenges are solvable (see SI). Therefore, with these improvements, USF will be a powerful fluorescence microscopic imaging technology with high SNR and high sensitivity for centimeter-deep tissues and can be used for many biomedical applications.

**Conclusion**

We have synthesized and characterized a NIR and environment extremely sensitive fluorophore, $ADP(CA)_2$, and a family of USF contrast agents based on this dye. These USF contrast agents have very high fluorescence intensity on-to-off ratios. A USF imaging system was developed to sensitively detect USF photons and efficiently suppress background noise. Also, a correlation algorithm was adopted to uniquely differentiate USF signal from background noise and differentiate one type of USF contrast agent from another one.

Based on these unique technologies, we have successfully achieved fluorescence microscopic imaging in centimeter-deep tissues with high SNR and picomole sensitivity using a laser and an ultrasound pulse below the safety thresholds. Multiplex USF fluorescence imaging is also demonstrated. This will be very useful in simultaneously imaging multiple targets and their interactions in the future, which are usually difficult for non-optical technologies. Also, this technology is demonstrated for the first time in both a tissue-mimic phantom and real biological tissues (porcine muscle tissue, *ex vivo* and *in vivo* mouse tissues).

**Methods.**

**Characterization of the dye and the USF contrast agents.** The adopted materials and the synthesis protocols of the dye and the USF contrast agents have been provided in SI. To measure the dye environmental sensitivity, a stock solution was prepared by dissolving ADP(CA)$_2$ in methanol at a concentration of 20 mg/mL (21.5 mM). For each experiment, a small portion of this stock solution was taken and dissolved it into different solvents. (1) *Polarity:* the five solvents with different polarity indexes were selected[19]. (2) *Viscosity*: eight solvents with different viscosities were prepared by mixing glycerol and ethylene glycol at different volume ratios: 0/100, 8/92, 16/84, 25/75, 50/50, 75/25, 90/10, and 100/0. The final concentration of ADP(CA)$_2$ in each solvent was 8.6 nM. The fluorescence measurement system was similar to the one used in our previous report (Ex: a 655-nm laser; Em: a 711/25-nm band-pass emission filter)[14].

**USF imaging system.** See the details in the section of "Setup of the USF imaging system" in SI.

**Definition of SNR of a USF image or profile.** In this study, the SNR of a USF image (or a profile) was defined and calculated based on the following rules. First, the USF profile along the x direction at each y location on a USF image was plotted. The peak strength of the profile was found and defined as signal strength (S). Second, a range on the x-axis centered at the peak location of the profile and with a width of three times of the FWHM of the profile was defined as the region where USF signal were potentially detected. Outside this region was defined as the background region. The standard deviation of all the data measured in the background region was defined as noise strength (N). Third, the ratio of signal strength to noise strength (S/N) was defined as the SNR of the profile. Fourth, the average and the standard deviation of the SNRs of all the profiles on each USF image were defined as the SNR of the USF image and the corresponding error.

**Multicolor USF, DF, and US imaging in silicone phantom.** The tissue-mimic phantom was made using silicone (to mimic tissue's acoustic properties) adopted with $TiO_2$ (to mimic tissue's optical scattering properties)[16]. The two edged micro-tubes were used for image co-registration and were filled with the corresponding contrast agent of each modality: air for US imaging; ICG-based agent for color-1 USF imaging; $ADP(CA)_2$-based agent for color-2 USF imaging; $ADP(CA)_2$-based agent for DF imaging. The procedures of DF imaging and ultrasound imaging were similar to those adopted in our previous paper[9].

**USF imaging in porcine muscle tissues.** The tissues were purchased from a local grocery store and cut with different thickness. The experimental parameters for the four thickness (0.8, 1.2, 2.2, and 3.1 cm) were as follows: laser power: 2.5, 2.5, 12.7, and 38.64 mW; corresponding laser intensity right before entering the tissue: 0.21, 0.21, 1.05, and 3.19 mW/cm$^2$; LIA sensitivity: 500, 500, 200, and 200 mV nA; HIFU driving voltage: 80, 80, 80, and 100 mV (the estimated mechanical index is 1.7, 1.7, 1.7 and 2.19). LIA time constant in all the experiments was set as 300 ms. All the experimental parameters used in Figure 3 were the same as those in the 8-mm thick tissue experiment described here.

**Bio-distribution and USF imaging in *ex vivo* mouse organs.** BALB/c mice (female, 20–25 gram) were purchased from Taconic Farms Inc. (Germantown, NY, USA). The animal protocols were approved by the University of Texas at Arlington's Animal Care and Use Committee. Animals were intravenously administered 100 μL final solution of the $ADP(CA)_2$-based contrast agents. Twenty-four hours later, the animals were sacrificed and their organs were rapidly dissected. The isolated organs were then immediately imaged using the In-Vivo FX Pro system (f-stop: 2.5, excitation/emission: 630/700 nm, 4×4 binning; Carestream Health, Rochester, NY, USA). The organs were then placed on the USF system for imaging. In this *ex vivo* USF study, all the USF-related parameters were identical to those in the 12-mm thick tissue sample USF imaging.

**Detection of *in vivo* USF signal from a mouse.** FVB/N mice (female, 20–25 gram) were purchased from Jackson Lab (Bar Harbor, ME, USA). All animals were anesthetized with 2%–2.5% isoflurane. Hair on the back and abdomen was removed, and 100 μL final solution of the ADP(CA)$_2$-based contrast agents were injected through a tail vein. After two hours, *in vivo* USF signal was acquired. All USF system parameters were the same as those in the 12-mm thick tissue sample USF imaging, except that the HIFU driving voltage was changed from 80 mV to 400 mV due to the significant dilution of the fluorophore concentration after injecting into a large volume (animal body) and possible thermo loss due to blood perfusion in a live animal.

**Acknowledgement**

This work was supported in part by funding from the NSF CBET-1253199 (BY), the NHARP 13310 (BY), the CPRIT RP120052 (BY), the NIH/NIBIB 7R15EB012312-02 (BY), the NIH/NHLBI R01 HL118498 (KN), and the NSF 1401188 (FD). The authors are grateful to Dr. Hong Weng for helps in bio-distribution measurement.

**Author Contributions**

B.Y. was the principal investigator and conceived the idea. B.C. and B.Y. developed the idea and designed the experiments. B.C. was the main operator of the experiments. B.C. and M.-Y.W. synthesized and characterized the contrast agents supervised by Y.H. and K.N.. B.C., Y.P. and B.Y. designed and implemented the imaging systems. B.C. conducted the imaging experiments, programmed signal processing algorithm and analyzed the data. V.B. synthesized and characterized the dyes supervised by F.D.. L.T. supervised the contrast agent bio-distribution experiment. B.C. and B.Y. explained the data and prepared the manuscript. All authors reviewed the manuscript.

**Additional information**

Supplementary information accompanies this paper.

Competing financial interests: The authors declare no competing financial interests.

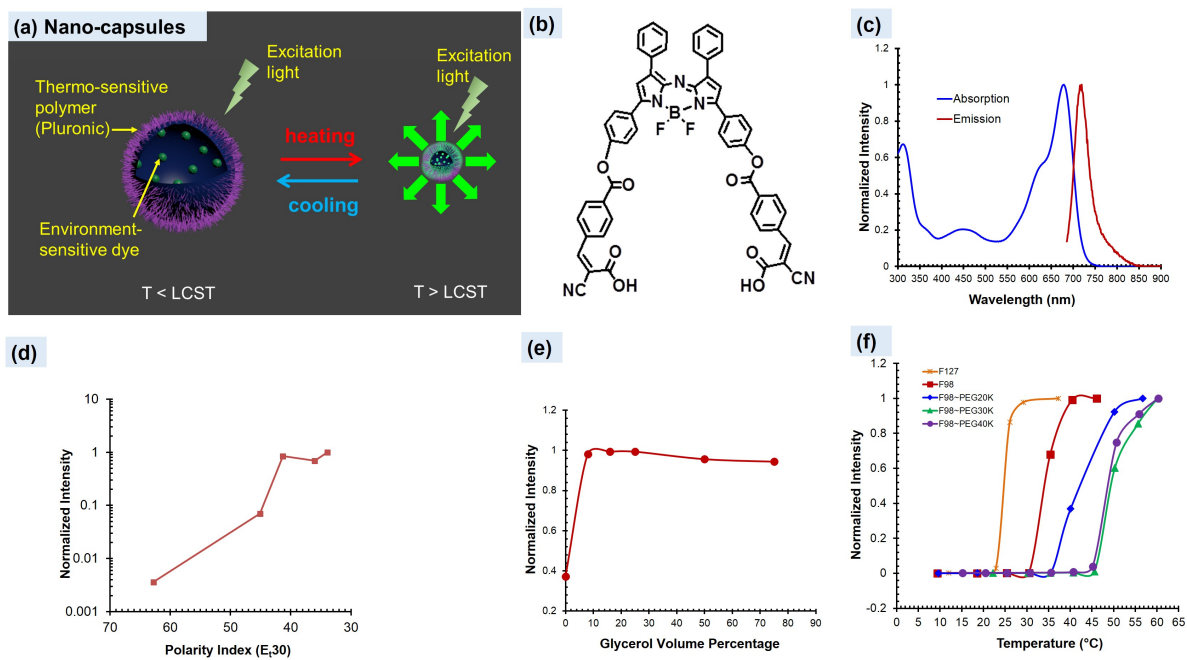

**Figure 1** Characterization of ADP(CA)$_2$ and ADP(CA)$_2$-based USF contrast agents. (a) A scheme displays the principle of USF contrast agents (LCST: lower critical solution temperature). (b) and (c) show the chemical structure, absorption and emission spectra (in dichloromethane) of the environment-sensitive dye, ADP(CA)$_2$. The fluorescence intensity of ADP(CA)$_2$ as function of polarity (d) and viscosity (e) of the solvent. Five solvents with different polarity index were employed, which are water (62.8), dimethyl sulfoxide (45.1), 1,2-dichloroethane (41.3), 1,4-dioxane (36) and toluene (33.9). A small polarity index represents low polarity. Viscosity was adjusted by varying the volume ratio of glycerol/(ethylene glycol). A high ratio means a high viscosity. (f) The switching relationship between the fluorescence intensity of these ADP(CA)$_2$-based USF contrast agents and the temperature: Pluronic-F127 (stars); Pluronic-F98 (squares); Pluronic-F98~PEG20k (diamonds); Pluronic-F98~PEG30k (triangles); Pluronic-F98~PEG40k (solid circles).

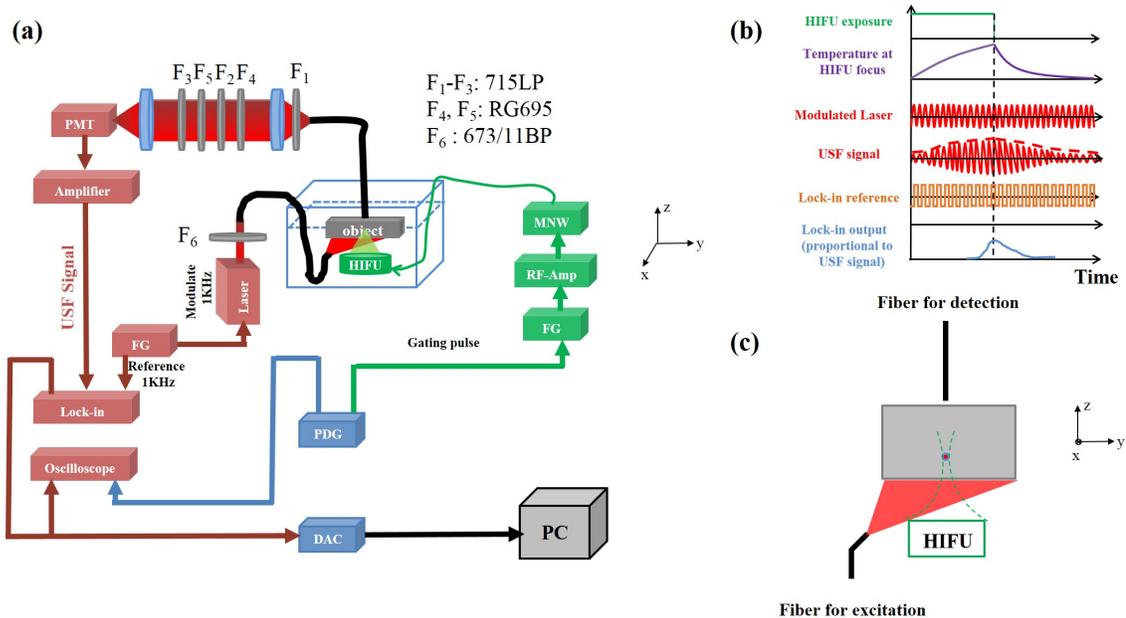

**Figure 2** The USF imaging system. (a) The schematic diagram of the USF imaging system. PDG: pulse delay generator; FG: function generator; RF-Amp: radio-frequency power amplifier; MNW: matching network; HIFU: high intensity focused ultrasound; PMT: photomultiplier tube; F1-F5: emission filters; F6: excitation filter; DAC: data acquisition card; PC: personal computer. (b) The time sequences of six different events in USF imaging, including HIFU transducer gating pulse, temperature change at HIFU focus, modulated laser output, USF signal, lock-in reference and lock-in output. (c) The sample configuration, including the sample, the excitation and emission fiber bundles, and the HIFU transducer.

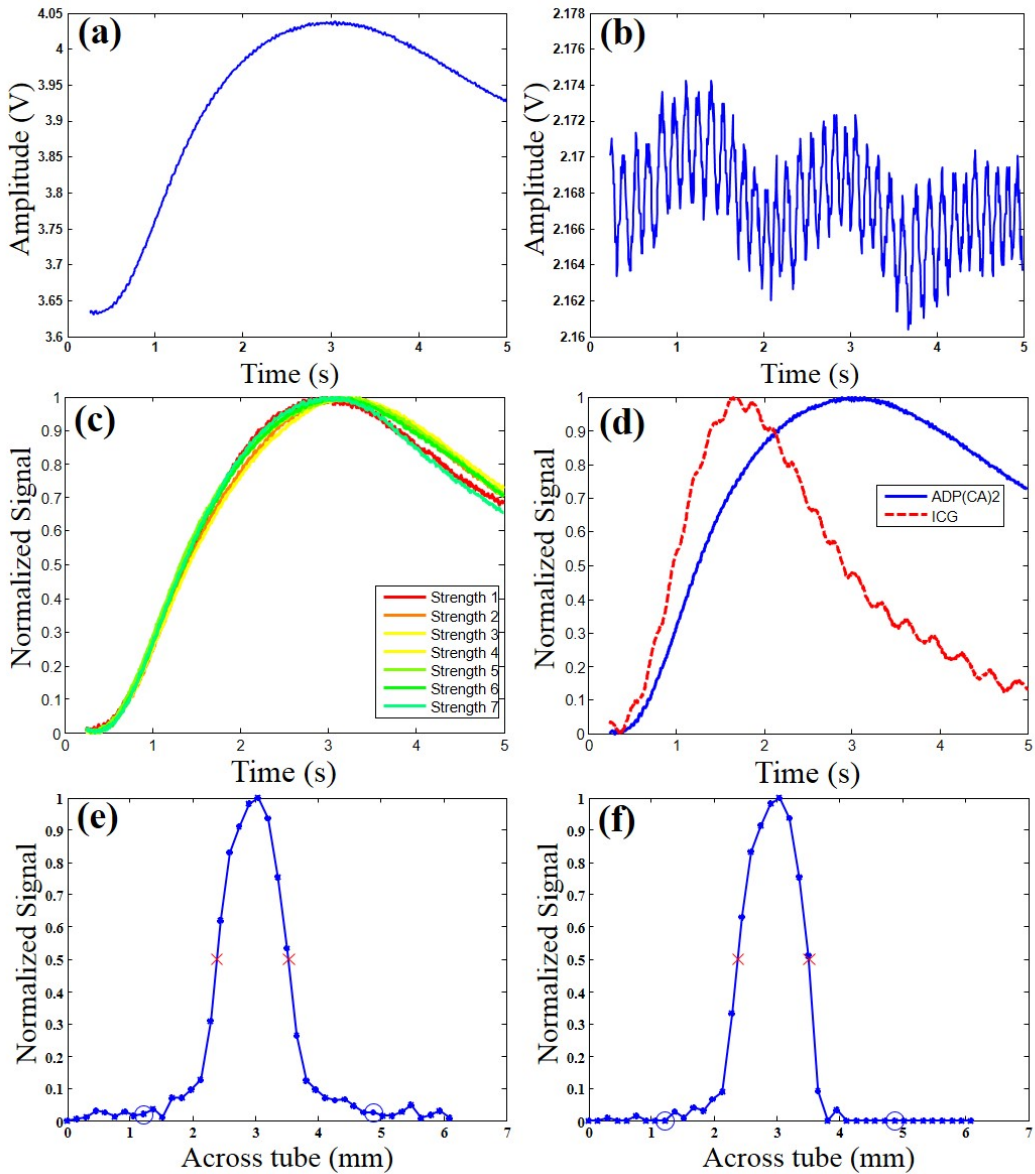

**Figure 3** Typical USF signals and the correlation method. Typical USF signal (a) and background noise (b) acquired from ADP(CA)$_2$-based USF contrast agents in the 8 mm-thick tissue USF experiment. (c) The normalized USF signal (from ADP(CA)$_2$-based agents) with different signal strengths (strength 1-7: 171, 239, 290, 310, 325, 306, and 270 mV, respectively) to show that the shape is independent of the signal strength. (d) The normalized USF signals from ADP(CA)$_2$-based (blue solid line) and ICG-based (red dash line) agents to show that the shape is dependent on the type of the agents. (e) and (f) are the normalized USF profile of the micro-tube (filled with the ADP(CA)$_2$-based agent) before (SNR: 88) and after (SNR: 300) correlation analysis, respectively. The experiment conditions are presented in Methods.

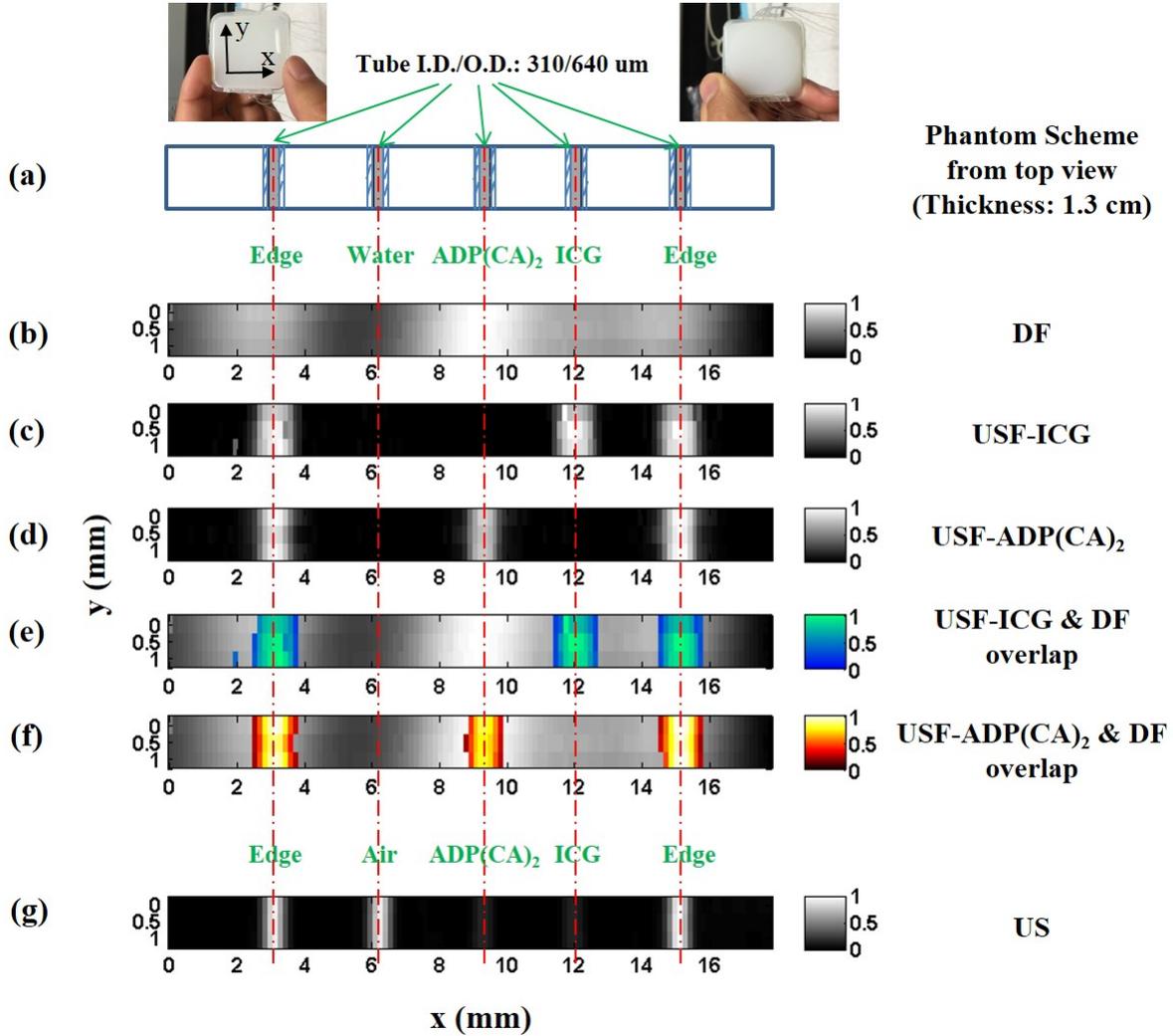

**Figure 4** Images of micro-tubes in a tissue-mimic phantom. (a) The photographs and schematic diagram of the tissue-mimic silicone phantom (on x-y plane): front view (left photo), back view (right photo) and the cross section of the micro-tube structure on the x-y plane (the bottom figure in (a)). (b) The image acquired by directly detecting fluorescence (DF) without ultrasound. (c) and (d) show the USF images acquired from ICG-based (color-1) and ADP(CA)$_2$-based (color-2) agents, respectively. (e) and (f) are the overlapped images between USF images ((c) and (d)) and direct fluorescence image (b). (g) Ultrasound image of the micro-tubes acquired from the same ultrasound transducer (i.e. a C-mode ultrasound image). The three non-edge tubes were filled with water (as background control), ICG-based agent (color-1) and ADP(CA)$_2$-based agent (color-2). The two edged micro-tubes were used for image co-registration and were filled with the corresponding contrast agent of each modality (see Methods).

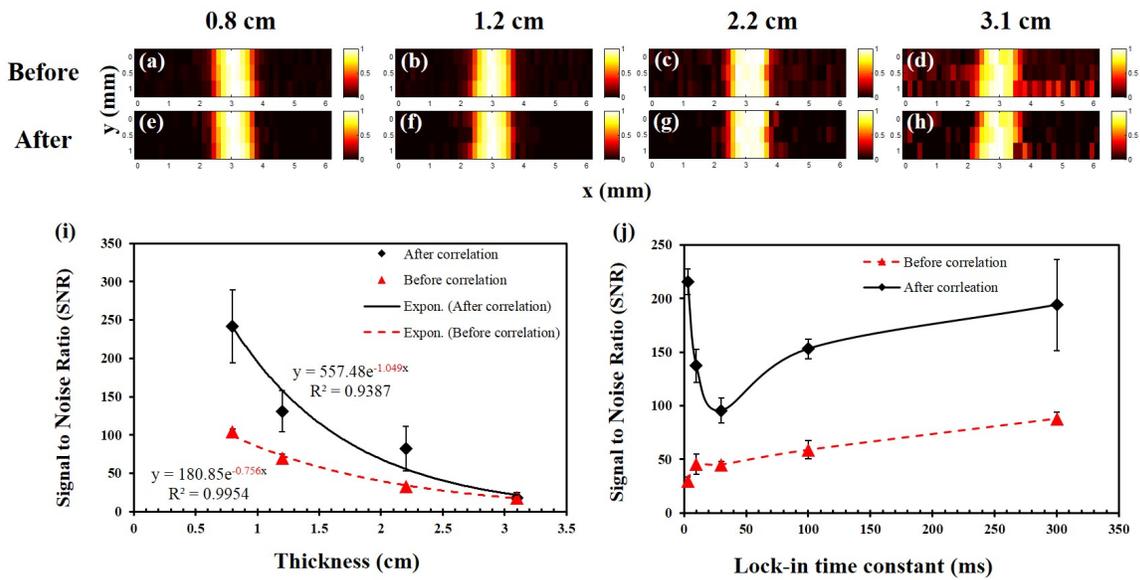

**Figure 5** USF images of micro-tubes in porcine muscle tissue samples. (a)-(d) the USF images of the micro-tubes (before correlation) that were embedded into pork muscle tissue samples with thicknesses of 0.8, 1.2, 2.2 and 3.1 cm, respectively. The I.D./O.D. of the tube is 0.31/0.64 mm; (e)-(h) The corresponding USF images processed by the correlation algorithm. (i) The relationship between the SNR and the thickness of the sample before (triangles) and after (diamonds) the correlation processing (error bar: mean±standard deviation). (j) The relationship between the SNR and the LIA time constant before (triangles) and after (diamonds) the correlation processing (error bar: mean±standard deviation).

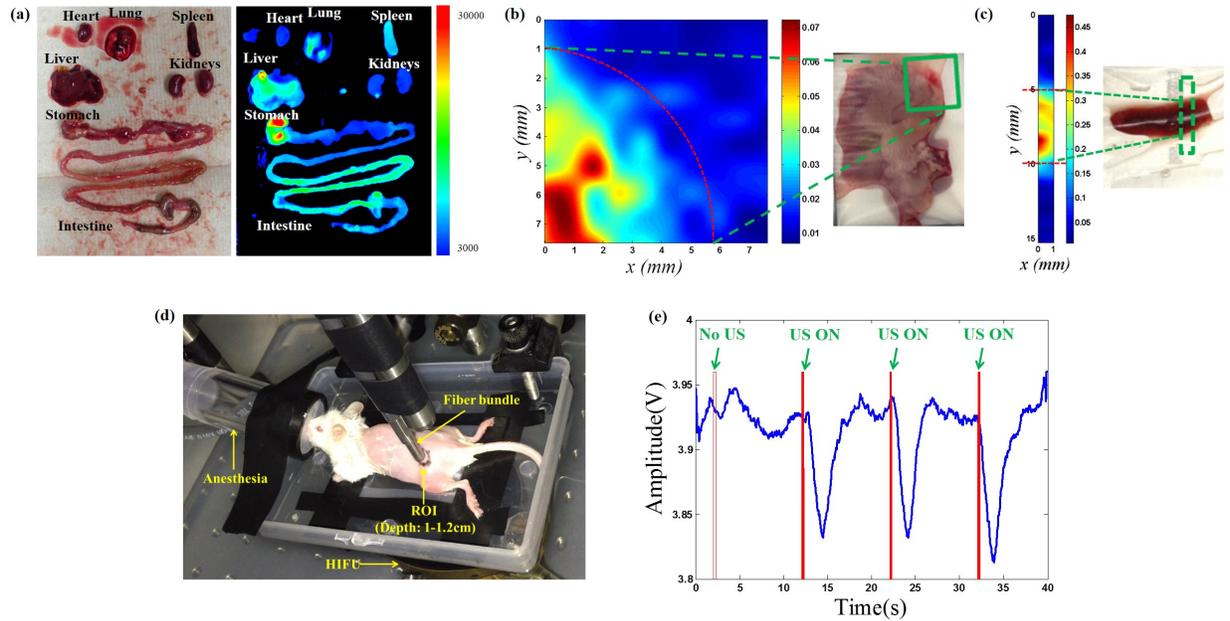

**Figure 6** USF images of *ex vivo* mouse organs and USF signals from *in vivo* mouse tissue. (a) Bio-distribution of ADP(CA)$_2$-encapsulated Pluronic-F127 nano-capsules in mice: a white-light photo (left) and the corresponding fluorescence image with Ex/Em=630/700 nm (right). (b) A USF image and a white-light photo of an *ex vivo* mouse liver. (c) A USF image and a white-light photo of an *ex vivo* mouse spleen. (d) A photo to show the basic experiment setup of *in vivo* USF imaging. (e) The acquired *in vivo* USF signal (the blue line) and the corresponding ultrasound exposure (the three sold red lines). The open red line represents no ultrasound pulse is applied.